# A short review on Applications of Deep learning for Cyber security

Mohammed Harun Babu R, Vinayakumar R, Soman KP

Center for Computational Engineering and Networking (CEN), Amrita School of Engineering, Coimbatore, Amrita Vishwa Vidyapeetham, India

Email: vinayakumarr77@gmail,com

**Abstract**

Deep learning is an advanced model of traditional machine learning. This has the capability to extract optimal feature representation from raw input samples. This has been applied towards various use cases in cyber security such as intrusion detection, malware classification, android malware detection, spam and phishing detection and binary analysis. This paper outlines the survey of all the works related to deep learning based solutions for various cyber security use cases. Keywords: Deep learning, intrusion detection, malware detection, Android malware detection, spam & phishing detection, traffic analysis, binary analysis.

## 1 Introduction

Cyber security involves protective key data and devices from cyber threats. It's a vital part of corporations that collect and maintain large databases of client data, social platforms wherever personal information were submitted and also the government organizations wherever secret, political and defense information comes into measure. It helps in protecting against vulnerable attacks that possess threat to special data, might or not across numerous applications, networks and devices. With the quantity of individuals accessing the information online which is increasing daily and also the threats to the data are increasing, with the cost of on-line crimes calculable in billions. Cyber security is that the set of technologies and processes designed to shield computers, networks, programs, and data from attack, unauthorized access, change, or destruction. These systems are composed of network security and host security systems, every of those has a minimum firewall, antivirus computer code, associated an intrusion detection system (IDS). This survey summarizes the importance of cyber security using Deep learning techniques (DL). Deep learning technique is been employed by researchers in recent days. Deep learning may be used along side the prevailing automation ways like rule and heuristics based mostly and machine learning techniques. This study helps is understand the advantage of deep learning algorithms to classify and correlate malicious activities that perceived from the varied sources like DNS, email, URLs etc. Not like ancient machine learning approaches, deep learning algorithms don't follow any feature engineering and have illustration ways. They will extract best options by themselves. Still, further domain level options got to outline for deep learning ways in information science tasks. The cyber security events thought-about during this study are enclosed by texts. To convert text to real valued vectors, numerous linguistic communication process and text mining ways are incorporated along with deep learning.

**2 Shared tasks**

In recent days, to enhance the system performance, shared task is organized as part of the conferences. In this shared task initially the train data set will be distributed among the participants and the train model will be evaluated on the test data set. This is most familiar in NLP domain recently shared task on identifying phishing email has been organized by [46] . The details of the submitted runs are available in [47]. Followed by shared task on detecting malicious domain organized by [48] as part of SSCC'18 and ICACCI'18. These two shared tasks allowed participants to share their approach through working notes or system description paper. Yearly there is one more shared task conducted by CDMC. But they don't have an option to submit system description papers. But this year (2018) they are giving an option to submit system description papers (CDMC 2018).

One significant issue was that the most of the publically available data sets are very old and each data set has their own limitations. To overcome such issues a brief study made to understand the need of Security domain, datasets and key feature of data sciences is discussed in [1] for problems employing the data science towards cyber security.

**3. Use-cases in cyber security**

**3.1. Intrusion detection**

An intrusion detection system (IDS) has been developed that's capable of detection every kind of network attacks within the environments. IDS detect malicious network activities by analyzing the collected packets, alarms to computer user, and blocks attack connections from attacks [2]. It additionally connects with the firewall as an elementary technology for network security. The characteristics of network is been assessed into Host based intrusion detection (HIDS) which involves in putting package and monitors internal packets of the system. To perform intrusion detection HIDS gathers data from its system calls, OS audit trails, application logs, etc. Network based intrusion detection which detects the malicious activity in network traffic. Generally, intrusion detection algorithms unit classified into 2 methods: misuse detection (Signature based) and anomaly detection. Signature based IDS: It is a technique looks a series of bytes or sequence with malicious network and helps in track down the detail log of the system which cause false alarm. Anomaly based IDS: It helps in identifying the anomalies and indicates serious and rare events overt the system and rectifies the unusual traffic pattern in a network. To resolve the disadvantage of these two detection method Hybrid IDS has been proposed which combines the complexity of anomaly and issue detection system and gets with new framework. Now-a-days, self-learning system becomes one of the prominent methods. Machine learning is one of the powerful concepts. Most of the ML solutions resulted in achieving the high false positive rate and high machine computation. This is due to most of machine learning techniques comes with the learning patterns among small-scale, low-level feature patterns of traditional and attack connections records. Most notably machine learning comes with deep learning which will be outlined as a better model of machine learning algorithms. These will help in learning the representation techniques with high advanced hierarchic sequence. In [3] proposed a model for novel deep learning approach in NIDS operation over networks, with combination of deep and shallow learning methods. This helps in analysis of network traffic over non symmetric deep auto encoder technique (NDAE). In [4], [5], [6] a brief study explains that Long short term memory(LSTM), Recurrent neural network(RNN), Convolution neural network(CNN) performs well in IDS systems when compared to other machine learning algorithms. CNN with n-gram technique is briefly discussed along with hybrid network such as CNN, CNN-recurrent neural network (CNN-RNN), CNN-long short-term memory (CNNLSTM) and CNN-gated recurrent unit (GRU).These techniques helps in identification of good and bad network ID in network connections. CNN has the capacity to gain high level feature representation from low level feature sets during extraction process is disused in [4]. Following with [5], [7] system call modeling based approach with ensemble method is proposed using LSTM algorithm for anomaly based IDS system. System call modeling helps in capture of semantic meaning of every call and relation over the network. Ensemble methods focus on false alarm rate which fits IDS design. This is a compact method, which helps in storage of parameters in a small space. This method is considered as fast and efficient approach in sequential matrix application. Application of deep neural networks is leveraged for intrusion detection by [52]. Recently, [54] discussed the various security issues in autonomous vehicles.

### 3.2. Malware detection

Malwares are programs which disrupts the data, files in the system which reduces the vulnerability and performance. In some cases it will lead to total corruption of system or a server [8]. These are easily passed through various environments using unauthorized software tools. There is existing some works in which deep learning has become one of the prominent methods in malware analysis [9]. In [10] binary and multiclassifier techniques are used for classification which gives better result when processed with rectified linear unit activation functions and dropout over the hidden networks. In [8], [11] deep learning approach applied with four layer network design is discussed. To get modest computation feature text extraction techniques such as Byte/Entropy Histogram Features, PE Import Features, String 2D histogram features; PE Metadata Features can be used. A brief discussion is made to show how to prevent overfitting and how backpropagation method helps in speeding up the learning process over the network. In [12], [9] echo state networks (ESNs) and recurrent neural networks (RNNs) helps in extracting full information by random temporal projection technique. Max pooling is used for non-linear sampling of data and logistic regression for final classification of data. In [13], discussed an advances malware technique known as Ransomware. It is a kind of crypto viral extortion which helps encrypt the files and gather information without other knowledge. In [14], [13] deep learning algorithm LSTM is been applied on API calls by binary sequence classification method. In [51] evaluated the performance of classical machine learning classifiers and deep neural networks on malware detection.

### 3.3. Android malware detection

Android device has becoming a popular nowadays among peoples. Malware detection becomes a big challenge in android platform. Deep learning along with NLP comes with a great breakthrough in this area [15], [16] Droid detector is a Google app helps in collect malware data. The collected data is processed for both static and dynamic analysis for feature extraction and it is characterized by DBN based approach. In [17] comes up with semantic information extraction from system call sequence method using NLP which helps in construction of deep learning model. LSTM model is constructed with effective number of hidden layers to achieve better result. Time cost function is used for classification by implementing different framework like Tensorflow [49] to speed up the process. This model is been compared with n-gram model which is considered as superior detection method in android malware. Hyper parametric tuning is been done in LSTM network, LSTM-RNN network topology explains how the architecture helps in get better result [18].[19], [20]. The effectiveness of the API call sequence is been studied, To perform this CNN is been approached by discussing the training size and sequence length which gives better indication in false and negative positive.

## 3.4. Detection and categorization of domain names generated by Domain name generation algorithms (DGAs)

Domain fluxing malwares are possessed through domain generation algorithm(DGA).These malwares encodes through domain or IP address by blocking the network from further communication to server and the host [21]. The detailed study on DNS log collection and deep learning for detecting malicious domain names in large scale is discussed in [22], [65]. In [23] explains DNS logs in side LAN environment which use deep learning algorithms for detection of malicious domain names and compared with the traditional machine learning algorithm. They claimed that the deep learning algorithms performed well in comparison to the traditional machine learning algorithms and moreover these algorithms remain as robust in an adversarial environment. In [24] describes a detail study on statistical feature approach on DGA systems by splitting the features into domain length, domain level using n-gram technique. In this approach Hidden markov model (HMM) is been used for classification. These traditional techniques are very slow and poor in the performance of false and true positives [25], [26]. [27] Deep learning technique helps in discrimination of DGA domains from non DGA domains. In [28] focused mainly on Character based method using neural networks such as RNN, CNN and Hybrid CNN. In RNN Endgame model is used which improves the model performance by adding dropout to overcome dropout during training phase along with embedding technique. To get a better predictive accuracy CMU model is implemented along with Bidirectional RNN [29]. The NYU and Invincea models are discussed with CNN and hybrid architecture of CNN is explained along with MIT model. All this models consists of multiple layers and are termed as most extensive architectures. In [30], [31] LSTM network comes up with an advantage of featureless extraction of raw domain names as an input is also discussed. In [50] proposed a unique framework which correlated the data's of DNS, URL and Email to increase malicious activities detection rate.

### 3.5. Spam and phishing detection

The study shows that the spam email is the act of sending undesirable information or mass information in a substantial amount to some email accounts. It is a part of electronic spam including almost indistinguishable messages sent to different beneficiaries by email. Along with phishing other cyber-crime technique scams other personal information such as passwords, credit card details, bank accounts etc. These problems are rectified using deep leaning techniques with Natural language processing (NLP). In [32] represented the phishing techniques over mail using unbalanced dataset. Mainly in this various techniques such as term frequency-inverse document frequency (TF-IDF), Nonnegative Matrix Factorization (NMF), bag of words are discussed for feature extraction and also algorithms such as Random forest(RF), logistic regression, k-nearest neighbor, Multi nominal navies Bayes are used. In which LR and MLB comes with high metric performance. In [33] neural network approach is discussed by applying pearl script for feature extraction which helps to get dataset in a vector format. A comparative study is done on the extracted dataset using Traditional machine learning algorithms in which Decision tree (DT), and neural network approach performed well. In [34] discussed about the NLP feature extraction techniques using methods such as character level embedding and word embedding. A comparative study is made among Support vector machine (SVM) using character level and CNN using both character and word embedding techniques. In which CNN using word embedding gives a better result. In [35] showed a new LSTM approach in which dataset are considered as a hierarchical email architecture by considering it as sentences and words. Bidirectional LSTM is used for both cases which helps in compute the weight and estimates the phishing probability over the data during the network computation. In [36] neural network is used for classification of URL phishing it consists of three layer linear network which makes the topology very light and compact. Malicious threat over URL's is been analyzed by character sequence. Embedding technique is used with RNN and hybrid CNN networks which helps in studying how to develop a shelter for web page content analysis from malicious URL's with faster web page response. In [53] the application of CNN is leverage for image spam detection. In [55] discussed the application of CNN and CNN-LSTM for phishing URL detection and compared with bi-gram text representation. Recently, a shared task on phishing email detection was organized as part of CODASPY'18 conference and the detailed information is discussed in [56-64].

## 3.6. Traffic Analysis

In [37] the density and the volume of internet traffic is been increasing day by day. Identification of data flow through the network is considered as a major problem in traffic analysis. In [38] discuss traditional method using Artificial neural network and deep learning methods and result shows that in feature learning, unknown protocol identification this approach very well but it could give better adaptation in non-automation method in traditional method. Deep packet framework for extracting features automatically form network traffic using Deep learning method is proposed in [17] these packets help in handle sophisticated task like multi challenging, traffics etc. In [39], [40] proposed an architecture for Shallow and deep network for secure shell protocol. In [41] which RNN network helps to classify and model the tunnel SSH by modeling the time series feature to identify statistical information of the traffic flow.

## 3.7. Binary Analysis

Binary analysis is a powerful security analysis tool which looks into binary codes and finds the vulnerability issues with uncertainty deploying in free and open software. Static analysis can understand the pattern of the code to find vulnerabilities. Now-a-days automated analysis method is combined with deep learning method which has overcome the pattern based limitations [42]. In [43] discussed about that different problems faced in binary analysis and how it can be solved using neural networks. Many benchmark approaches are analyzed with RNN, LSTM and GRU networks. To rectify Gradient descent over the hidden layers, optimization is done with rmsprop method. This is also been discussed with background behind all networks with time function, error analysis, function identification and limitation of network. In [44] also came with RNN network but in this work so many experiments are performed on machine code snippets. These snippets are performed on LLVM and MIPS binaries data using tokenization scheme. These helps in extraction of higher level structure from lower level binary structure. In [45] presents a new system EKLAVYA which helps in recovering function type signatures from disassembled binary codes using RNN network. Argument recovery module on RNN is implemented using techniques like saliency mapping and sanitization. This system helps in learn calling conventions and idioms with high level accuracy parameter. In [42] discussed deep learning method on assembly codes which help to analysis the software weakness. In this Text-CNN is been analyzed using Instruction2vec and word2vec which has given high accuracy in classification of text data.

## 4. Conclusion

Deep learning is a prominent algorithm employed in several cyber security areas. Considering several traditional methods and machine learning methods deep learning algorithms considered as a robust way to solve problems. From this study it is clear that most of the deep learning algorithms comes up with better accuracy rate, which will be helpful in building an real time application for analyzing malicious activities over network.